\documentclass[final,3p,times]{elsarticle}
\usepackage{amssymb}
\usepackage{amsmath}

\journal{J Mech Phys Solids}

\begin{document}

\begin{frontmatter}

\author{Aoi Nohara}
\author{Ko Okumura}
\affiliation{Physics Department and Soft Matter Center, Ochanomizu University, 2-1-1\\
Ohtsuka, Bunkyo-ku, Tokyo 112-8610, Japan}

\title{Universality in the velocity jump in the crack propagation observed for
food-wrapping films for daily use}

\date{\today}

\begin{abstract}
The velocity jump found in the crack propagation for rubbers has been a
powerful tool for developing tough rubber materials. Although it is suggested
by a theory that the jump could be observed widely for viscoelastic materials,
the report on a clear jump is very limited and, even in such a case,
reproducibility is low, except for elastomers. Here, we use a mundane
food-wrapping film as a sample and observe the crack propagation velocity with
pulling the sample at a constant speed in the direction perpendicular to the
crack. As a result, we find the jump occurs at a critical strain with high
reproducibility. Remarkably, the plot of the crack-propagation velocity as a
function of strain can be collapsed onto a master curve by an appropriate
rescaling, where the master curve is found to be universal for change in the
pulling speed and in the sample height. The result reveals a key parameter for
the jump is the strain, suggesting the existence of a small length that
characterizes the deformation along the crack. The present study shows that
the jump could be observed if the material exhibits a glass transition in
rheology, even if more than one structural relaxation occur as in many
viscoelastic materials, although a single relaxation of the glass structure is
assumed in the previous theory. This finding suggests the possibility that the
velocity jump could be observed in any viscoelastic materials that exhibits a
glass transition. Our results set limitations on future theories and open an
avenue for the velocity jump to become a tool for developing a wide variety of
tough polymer-based materials.
\end{abstract}


\end{frontmatter}


\section{Introduction}

To develop tough polymer-based materials are crucial in our modern life.
Accordingly, the fracture of polymer based materials has been studied
intensively \cite%
{PGGtrumpet,Persson2005review,Creton2016review,saulnier2004adhesion,Hui2003,LakeThomas1967}%
, but remains an active area of research. One of the principal topics in the
study of polymer fracture is the crack propagation, which can be studied
through various methods: the cyclic tensile test \cite%
{Heinrich2002,Tanaka1974}, trouser test \cite{Gent1996Langmuir,Lindley1972},
static tensile test \cite{Tsunoda2000,MoridhitaUrayama2016PRE}, and so forth 
\cite{creton2013softmatter,Lin2010}.

Among a myriad of polymer-based materials, rubber is one of ancient matters
human beings have been exploited and is used on a daily base in various
forms from automobile tires to anti-vibration system for buildings. In the
study of fracture of rubber or elastomer, the velocity jump has been
discussed systematically \cite{Tsunoda2000,MoridhitaUrayama2016PRE} and more
recent years extensively with an increasing expectation for toughening \cite%
{morishita2017crack,morishita2019universal,kubo2017velocity,sakumichi2017exactly,Okumura2018}%
.

Conventionally, the jump is observed in experiments performed in a static
experiment performed under a static boundary condition. In the static
experiment, a film sample with the width much larger than the height is
stretched in the height direction, and with a fixed stretch a cut is
introduced at one of the free edges. The cut triggers crack propagation in
the direction of width and the propagation speed reaches a constant value,
which defines the crack-propagation velocity for the fixed strain. The
velocity as a function of strain increases as the strain increases, which
may be intuitively natural. Remarkably, the velocity jumps typically 10000
times at a critical strain. This dramatic phenomena is called the velocity
jump.

In 2017, a theoretical explanation was proposed based on a simple
viscoelastic model, which suggests the jump can be observed for a wide range
of viscoelastic materials \cite{sakumichi2017exactly}. The jump was
reproduced in numerical simulation \cite{kubo2017velocity}, and the physical
picture for the origin of the velocity jump proposed in Ref\cite%
{sakumichi2017exactly}, the glass transition at the crack tip, is
strengthened by a study performed under a collaboration among theoretical,
numerical and experimental groups \cite{kubo2021dynamic}.

Since the suggestion that the jump can be observed for viscoelastic
materials other than rubber, some attempts have been made to observe the
jump in other polymer-based materials, such as gel \cite%
{liu2019crack,zhang2022unique}. However, observation of a clear jump other
than elastomers has been limited. One such example is known for a porous
sheet made of polypropylene, but the reproducibility of the experiment was
low \cite{TomizawaJump2}. Although a clear jump, of the order of four orders
of magnitude, was observed for any samples used, the quantities
characterizing the jump, such as the critical strain and the velocities just
before and after the jump, were clearly dependent on samples.

As a matter of fact, the velocity jump was not observed for the porous
polypropylene, when the crack propagation was observed under the static
boundary condition \cite{Takei2018}. However, the jump was observed in a
dynamic experiment, in which the crack propagation was observed under a
dynamic boundary condition \cite{TomizawaJump2}: we first introduce a short
crack at one edge of the sample, and then start stretching the sample at a
fixed speed to observe the crack propagation \cite{AoyanagiDynamic}. The
reason the jump was not observed under the static experiment may be that the
sample is relaxed during the preparation for giving a fixed stretch, while
such effects of relaxation can be reduced in the dynamic experiment. This
higher sensitivity for detecting the velocity jump is one of the advantages
of the dynamic experiment over the static one.

Another advantage of the dynamic experiment is efficiency in time and cost.
In the static experiment, we need many samples to obtain a velocity curve as
a function of strain, because we need to break one sheet to obtain a single
point on the curve. On the contrary, in the dynamic experiment, we can
obtain a full curve with a single sheet: we can measure the velocity as a
function of strain during crack propagation because the strain changes as we
stretch the sample.

Here, we use a mundane food-wrapping film as a sample and observe the crack
propagation under the dynamic boundary condition. As a result, we find the
jump occurs at a critical strain with high reproducibility. The crack-tip
position as a function of time is obtained using two cameras, one for
capturing the slow crack propagation before jump and the other with a high
frame rate for the fast crack propagation after jump. The plot of the
crack-propagation velocity as a function of strain thus obtained shows a
little sample dependence: for a fixed pulling velocity and a fixed sample
size, we obtained quantitatively similar curves for different samples.
Furthermore, we find that the velocity curve is independent of the pulling
velocity but dependent on the sample height, while they can be collapsed
onto a master curve by rescaling: the master curve thus obtained is
universal for change in the pulling speed and the height of the sample. The
result reveals a key parameter for the jump is the strain, suggesting the
existence of a small length that governs the deformation along the crack.
The present study sets limitations on future theories and opens an avenue
for the velocity jump to become a tool for developing a wide variety of
tough polymer-based materials important for our life. For example, the
present study will be relevant for resolving one of urgent issues in modern
society, the fire accident of mobile battery by toughening a polymer sheet
called the separator in secondary batteries, the deterioration of which
leads to the fire accident. Note that the separator is a polymer sheet
similar to the sample used in the present study in the two respects: both
sheets have similar thicknesses and both are made of crystalline polymers.


\section{Materials}

The sample we used in the present study is a commercially available product,
SARAN-WRAP (Asahi Kasei Corp.). This film is transparent and sold for use in
daily life to wrap food for a short preservation. The film of thickness 11 $%
\mu $m is made mainly of polyvinylidene chloride (PVDC) and sold as a roll
of width $W=15$ cm and of length 50 m. According to the patent \cite%
{AsahiKaseiPAT}, the film is fabricated under the inflation process, with
anisotropic extension: the extension ratio in the direction of the material
flow, called the machine direction (MD) is typically three times, while the
ratio in the perpendicular direction, called the transverse direction (TD),
is larger, typically five times. The anisotropic extension leads to a high
degree of crystalline orientation in the TD direction, which is typically 85
per cent. The tearing strength in the MD direction (i.e., when the teared
crack runs in the MD direction) is higher than the tearing strength in the
TD direction. The film is easy to cut in the TD direction, which corresponds
to the width direction of a roll, and strong for tearing in the MD direction.

From the strain-stress curve obtained at the room temperature at the strain
rate $\simeq 0.002$ to 0.07 s$^{-1}$, the elastic modulus is around 500 MPa
and increases with the strain rate. From rheological measurements, the
sample exhibits a glass transition at the time scale around tens of
milliseconds at the room temperature, where the elastic modulus at the glass
phase $\simeq 5$ GPa drops to a smaller value $\simeq 500$ MPa. See Appendix
for further details on mechanical and rheological properties.

\begin{figure}[ptb]
\begin{center}
\includegraphics[width=0.8\textwidth]{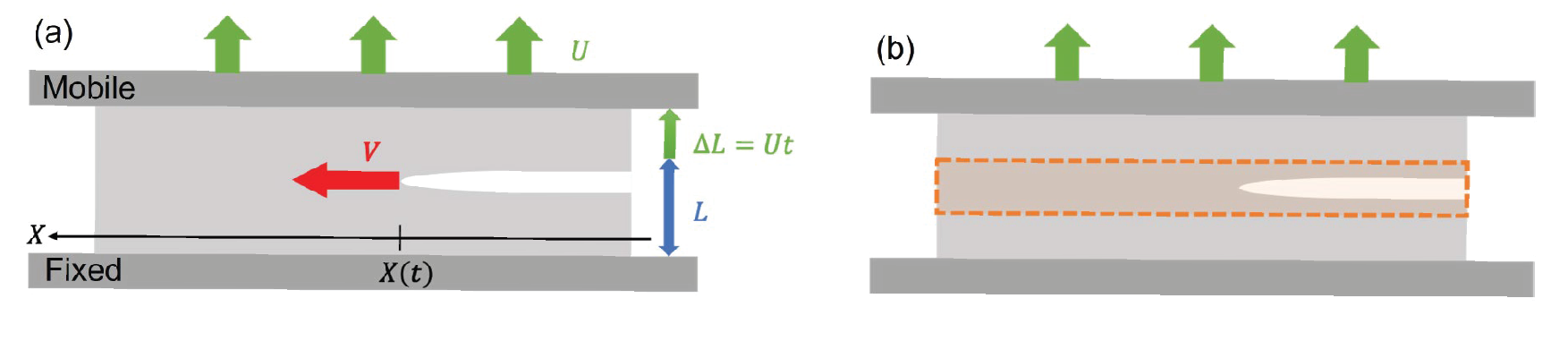}
\end{center}
\caption{(a) Experimental setup for the dynamic experiment. (b) Illustration
for the long and thin tape-like area along the crack path.}
\label{Fig1}
\end{figure}

\section{Experimental}

As illustrated in Fig.~\ref{Fig1}, we cut a sheet sample from a roll of
width $W$ ($=150$ mm) and set the sheet bottom to the fixed clamp and the
sheet top to the mobile clamp so that the sample height at the zero strain
becomes $L$ ($=10$ to 25 mm), i.e., the distance between the bottom and top
clamps is set to $L$ at the zero strain. We introduce a short line crack of
length $a$ at one of the free edge, before we start stretching the mobile
clamp at a constant speed $U$ ($=0.5$ to 2 mm/s). The mobile clamp is
connected to a slider (EZSM6D040K, Oriental Motor) controlled by a
controller unit (ESMC-K2, Oriental Motor) via a wire. The strain rate $U/L$
is in the range 0.025 to 0.2 s$^{-1}$ in the velocity jump experiment. All
experiments are performed at the room temperature $\simeq 25$ $^{\circ }$C.

Some of the stress-strain curve given in Appendix are obtained by the same
setup used for the velocity jump experiment. But, in this case, the mobile
clamp is not connected to the slider system directly via wire but to a
digital force gauge (FGP-50, NIDEC-SHIMPO) mounted on the slider system via
a wire.

We record the crack propagation using two cameras. In order to capture the
slow propagation before the jump, we record a video at 60 frames per second
(fps) with a digital camera (D800E, Nikon) with a lens (AF-S Micro NIKKOR
60mm 1:2.8G ED, Nikon). In addition, to capture the fast propagation after
the jump, we record images at 32000 fps with a high-speed camera (FASTCAM
Mini UX100, Photron) with a lens (AF-S NIKKOR 20mm 1:1.8G ED, Nikon).

To clearly observe the crack propagation, before we set the film sample to
the clamps, we paint the long and thin tape-like area along the crack path
as shown in Fig.~\ref{Fig1} (b), whose area is $Wl$ with $l$ typically 1 cm,
with an oil-based white marker (Mackee Paint Marker, Zebra). But, later, to
see the effect of painting, we performed experiments by painting the whole
area ($=WL$) with a different white water-based pigment marker (POSCA Extra
Thick Square Core, UNI). In the following, we mainly discuss the data
obtained from samples with the whole area painted and compare the data
obtained from partially painted films. In the two different experiments, we
also change the length of the initial crack $a$: in the partial painting
experiment we set $a=10$ mm, but in the whole painting experiment, we set $%
a=L$. By comparing the two experiments, we can gain insight into the effect
of both painting and the initial crack length $a$.

\begin{figure}[ptb]
\begin{center}
\includegraphics[width=0.6\textwidth]{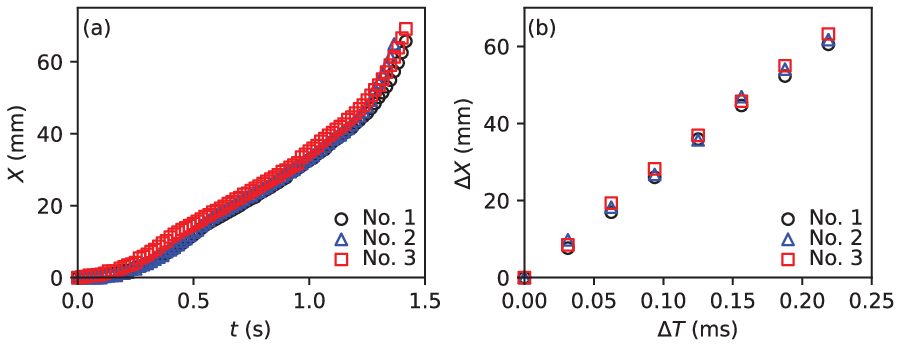}
\end{center}
\caption{(a) Crack tip position $X$ vs.~time $t$ before the jump at $X=X_{c}$
and $t=t_{c}$. (b) Crack tip position $\Delta X$ vs.~time $\Delta T$ after
the jump, where $\Delta X=X-X_{c}$ and $\Delta T=t-t_{c}$. The three sets of
the data are obtained from different three samples for $U=1$ mm/s and $L=15$
mm.}
\label{Fig2}
\end{figure}

\section{Crack propagation}

From the images obtained from the digital camera, we track the crack tip
position $X$ as a function of time $t$, before the jump, as in Fig.~\ref%
{Fig2} (a), where we showed the results from three samples obtained for the
same set of $(U,L)$. The three sets of the results are well overlapped,
demonstrating high reproducibility of the slow propagation. As for the fast
propagation after jump, we track the crack position $\Delta
X(t)=X(t)-X(t_{c})$ as a function of time $\Delta T=t-t_{c}$ with $t=t_{c}$
corresponding the time of jump, as in Fig.~\ref{Fig2} (b), where we show the
results from three samples obtained for the same set of $(U,L)$. The three
sets of the results are again well overlapped, demonstrating high
reproducibility even for the fast propagation.

\begin{figure}[tbp]
\includegraphics[width=\textwidth]{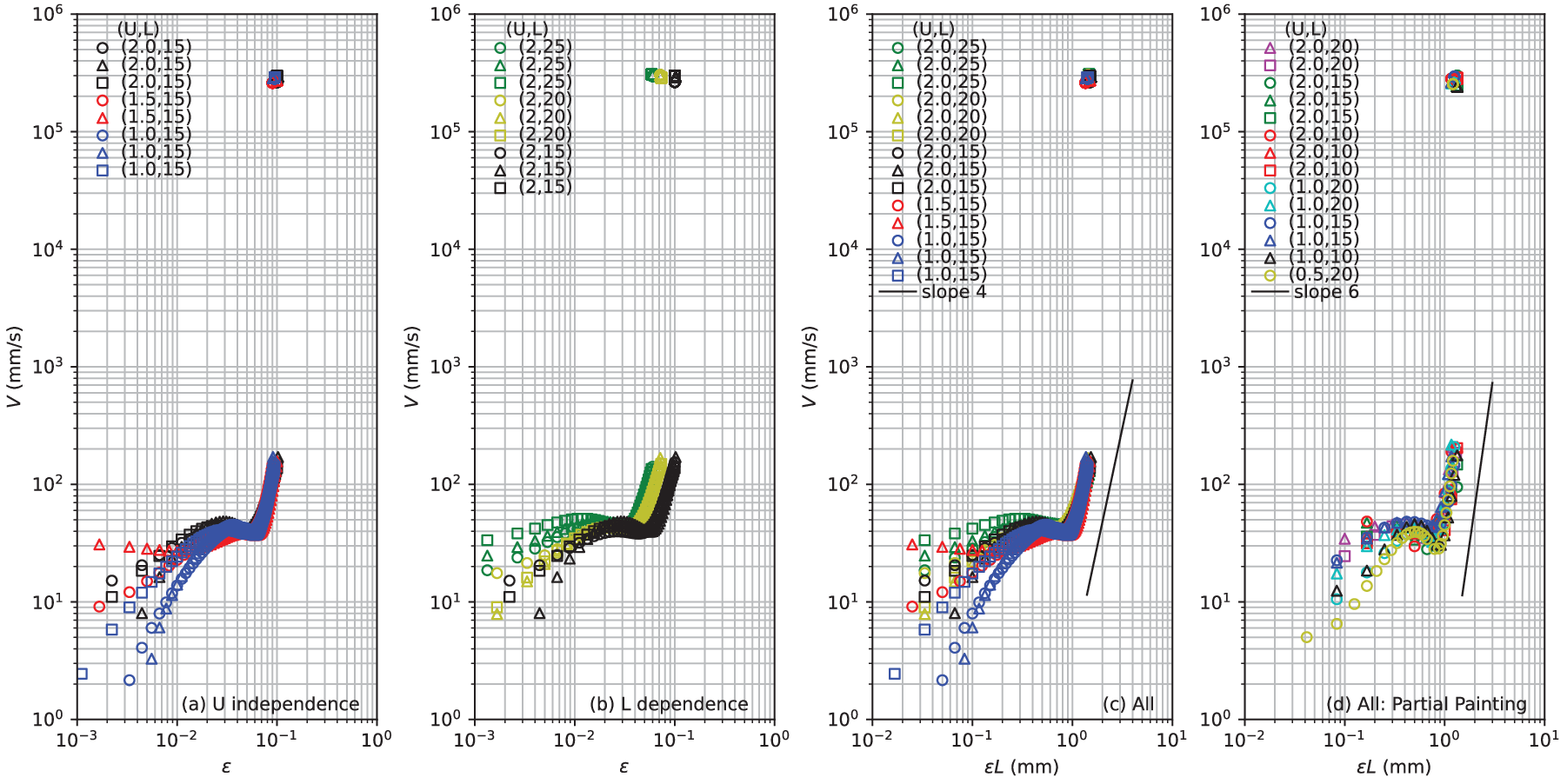}
\caption{(a) Velocity $V$ vs.~strain $\protect\varepsilon $ for a fixed $L$,
exhibiting a collapsed straight region on the log-log scale before the jump
and a collapsed point just after the jump. (b) $V$ vs.~$\protect\varepsilon $
for a fixed $U$. (c) All the data in (a) and (b) on a rescaled horizontal
axis $\protect\varepsilon L$, showing $V$ is a function of $\protect%
\varepsilon L$ in the straight region and at the points just before and
after the jump. (d) $V$ vs.~$\protect\varepsilon L$ obtained for various
parameters by a different way of painting to visualize the crack
propagation: partial painting (see the text for the details).}
\label{Fig3}
\end{figure}

\section{Velocity jump}

In Fig.~\ref{Fig3} (a), we show the crack-propagation velocity as a function
of strain for various pulling speed $U$ at a given sample height $L$ ($=15$
mm), which shows a clear jump. The velocity jump of three orders of
magnitude occurs at a critical strain $\varepsilon_{c}\simeq0.1$ with the
velocity just before the jump $V_{B}\simeq200$ mm/s and that just after the
jump $V_{A}\simeq300$ m/s. The curve shows a clear straight region in the
log-log plot before the jump. The straight region and the velocities just
before and after the jump, $V_{B}$ and $V_{A}$, are superimposed well even
though the data with different $U$ are shown, which demonstrates with a high
reproducibility the crack-propagation dynamics leading the jump is
independent of the pulling speed $U$.

In Fig.~\ref{Fig3} (b), we show the same relation for various sample height $%
L$ at a given pulling speed $U$ ($=2$ mm/s), which again shows a clear jump.
We see that the velocities just before and after the jump, $V_{B}$ and $V_{A}
$, are almost the same, while the dependence of the critical strain on $L$
is recognized.

In Fig.~\ref{Fig3} (c), we show all the data in Fig.~\ref{Fig3} (a) and (b)
with rescaling the horizontal axis from $\varepsilon$ to $\varepsilon L$.
The plot clearly demonstrates that all the data in the region from the
straight region to the point after the jump collapse onto a master curve,
which is universal for changes in $U$ and $L$. The slope $\alpha$ of the
straight region is $\sim4$.

To see the effect of painting and the initial crack length $a$, we show the
data obtained from partially painted films in Fig.~\ref{Fig3} (d), with a
straight line on the log-log plot with a slope $\alpha \sim 6$. Although we
see slight changes in $V_{B}$ and $V_{A}$ as well as in the slope of the
straight region from Fig.~\ref{Fig3} (c), the basic features are unchanged:
(1) $V=F(\varepsilon L)$ except for the small strain region with $%
F(\varepsilon L)\simeq (\varepsilon L)^{\alpha }$, where the master curve $%
F(x)$ is universal even if we change $U$, $L$, and $a$ in limited ranges,
although the slope $\alpha $ is affected by the way of painting (we changed $%
a$ in the whole painting experiment, which suggests the result is
independent of $a$). (2) $V_{B}$ and $V_{A}$ are also universal, i.e.,
almost independent of $U$, $L$, $a$, and the way of painting.

To obtain velocity of the slow crack propagation in the above analysis, we
fit the relation between the crack tip position $X$ and time $t$ with the
sixth-order polynomial of $t$ and take the derivative of the polynomial with
respect to $t$. We checked that the results do not exhibit significant
changes if we use the fifth-order polynomial except for the small strain
region before the straight region on the log-log plot. The velocity of the
fast crack propagation was obtained by a linear fitting.

\section{Discussion}

We find the crack propagating velocity is given as a function of $\Delta
L=\varepsilon L$, i.e., the amount of the absolute stretch $\Delta L$,
regardless of the sample height $L$. We interpret this unexpected result as
follows. There is a characteristic microscopic length scale $\simeq l^{\ast }
$ near the crack and practically only the region near the crack tip
characterized by $l^{\ast }$ is stretched. In other words, the strain near
the crack tip is not governed by $\Delta L/L$ but by $\varepsilon ^{\ast
}=\Delta L/l^{\ast }$ and the velocity is a function of this effective
strain $\varepsilon ^{\ast }$. One possible candidate for the scale $l^{\ast
}$ may be defined by the velocity just before the jump $V_{B}\simeq 200$
mm/s and the characteristic time scale for the glass transition $t_{0}$ ($%
\simeq $ tens of milliseconds) that we obtain from the rheology experiment: $%
l^{\ast }=V_{B}t_{0}\simeq $ a few mm. In this respect, the experiment with
smaller $L$ with $L<l^{\ast }$ is interesting, although technically
challenging.

The crack propagation velocity has been frequently plotted as a function of
the fracture energy $G$ \cite%
{GreenwoodJohnsonRate,Persson2005PRE,Langer1989PRA,Bouch2011}, which scales
as $\varepsilon^{2}L$ for liner elastic materials. This is the case in most
of previous studies on the velocity jump (e.g., Ref\cite%
{morishita2019universal}). However, we confirmed that, if we re-scale the
horizontal axis not to $\varepsilon L$ but to $G\sim\varepsilon^{2}L$ in
Fig.~\ref{Fig3} (c) and (d), the collapse to a master curve is not observed
anymore (they become scattered as in Fig.~\ref{Fig3} (b)). Note that, in the
present case, the critical strain at the jump is relatively small, up to
which $G$ can be obtained with a reasonable precision under the assumption
of a linear stress-strain curve (see Appendix).

The elastic modulus and the density of the present sample are $E\simeq 500$
MPa and $\rho \simeq 1000$ kg/m$^{3}$ at the room temperature. The
corresponding velocity of elastic wave is $v\simeq \sqrt{E/\rho }\simeq 500$
m/s, which gives a practical maximum of the crack propagation speed and is
comparable with the velocity after the jump $V_{A}\simeq 300$ m/s: After the
jump, the velocity reaches the maximum speed. This is in contrast with the
case of elastomers, where $V_{A}$ is a few orders of magnitude slower than
the velocity of the elastic wave $v\simeq 50$ m/s evaluated for the elastic
modulus of the rubbery regime $E\simeq $ a few MPa (e.g., Ref\cite%
{morishita2019universal}). Because of this, for elastomers we observe a
"fast-mode" in the velocity range from $V_{A}$ to $v$, in which $V$ scales
with some power of $\varepsilon $. However, it could be possible that we
observe velocity increase after the jump even in the present case if the
sample width is much longer, or we perform the static experiment with
minimizing the preparation time for relaxation in setting samples.

The analytical theory is based on a viscoelastic model characterized the
following parameters \cite{sakumichi2017exactly}: (1) a rubber elastic
modulus $E_{R}$ and a glassy one $E_{G}=\lambda E_{R}$ where $\lambda$ is
typically $10^{3}$ for elastomers, (2) the sample height $L$ and a
microscopic parameter $l_{0}=L/N$, (3) a viscosity $\eta$, and (4) a
critical strain $\varepsilon _{c}$ for rupture. The theory predicts the
normalized crack-velocity $V/V_{0}$ with $V_{0}\sim(l/\eta)\sqrt{E_{R}}$ as
a function of the normalized energy density $w/w_{0}$ with $w\sim
E_{R}\varepsilon^{2}$ and $w_{0}$ $\sim E_{R}\varepsilon_{c}^{2}/N$. Since $%
w/w_{0}\sim N\varepsilon^{2}/\varepsilon_{c}^{2}\sim\varepsilon^{2}L$ and
the fracture energy $G\sim\varepsilon^{2}L$, the theory suggests that $V$ is
given as a function of $G$ for a given material: $V=F(G)$. In addition, the
theory predicts $V_{B}\sim\sqrt{\lambda}V_{0}$ and $V_{A}\sim\sqrt{N}V_{0}$
as for the characteristic velocities and predicts the energy density at the
jump as $w_{j}\sim\lambda w_{0}$, which leads to a prediction for the strain
at the jump: $\varepsilon_{j}^{2}\sim(E_{G}/E_{R})/N$.

However, a close comparison of the theory with the present experiment is
difficult. This is because, while the theory is based on the single
relaxation from the glass phase to the rubber phase, the present sample
shows two relaxations: first for the glass phase and the second for the
crystalline phase (see Appendix). Our experiments suggest, even in such a
case, the glass transition, a common property in both cases of the theory
and the present experiment, is the key for the velocity jump.

\section{Conclusion}

We found a clear velocity jump in a highly reproducible manner for a polymer
film we use on a daily base under the dynamic experiment. The velocity -
strain relation can be collapsed onto a master curve after rescaling, which
is universal for the stretching speed and the sample size in ranges we
investigated. This suggests the jump is governed by the strain and there
exists a small length characterizing the deformation along the crack. The
rheological measurement suggests that the velocity jump could be observed
not only in polymer films with a single relaxation, which is assumed in the
previous theory, but also in films with multi-relaxations, if they exhibit a
glass transition. Considering that many viscoelastic materials shows a glass
transition with multi-relaxations, the present results underscore the
potential of the dynamic crack-propagation experiment for developing tough
polymers important for our life such as the separator in secondary batteries
and the need for a theory based on viscoelasticity with multi-relaxations.

\vspace{5pt} \textit{\ We thank Takashi Uneyama (Nagoya University) and
Naoyuki Sakumich (Tokyo University) for discussions respectively on
rheological properties of the sample and on general features of the
crack-propagation velocity. We are grateful to Anton Paar Japan K.K. for
helpful technical support concerning rheological measurements. This work was
supported by JSPS KAKENHI Grant Number JP19H01859 and JP24K00596 and a grant
from Eno Science Foundation. } 

\section*{Appendix: Mechanical Responses}

\begin{figure}[ptb]
\includegraphics[width=\textwidth]{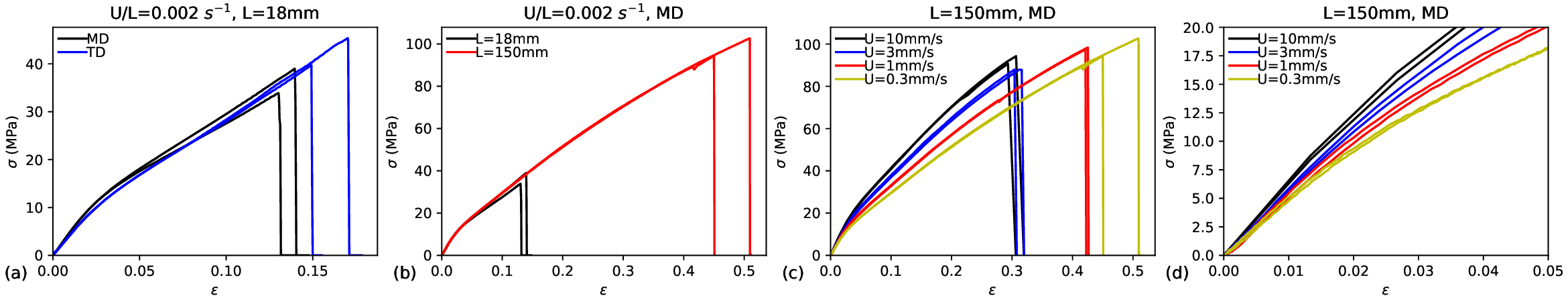}
\caption{Stress vs.~Strain. (a) Dependence on the stretching direction. (b)
Dependence on the sample size. (c) Dependence on the stretching speed. (d)
Magnified version of (c).}
\label{Fig4}
\end{figure}

\begin{figure}[ptb]
\begin{center}
\includegraphics[width=0.4\textwidth]{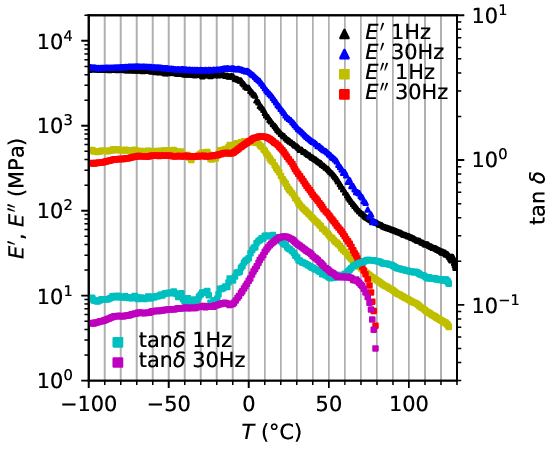}
\end{center}
\caption{Complex modulus and $\tan\protect\delta$ as a function of
temperature at $f=1$ and $30$ Hz.}
\label{Fig5}
\end{figure}

\subsection{Stress vs.~Strain}

In Fig.~\ref{Fig4}, we show the stress-strain curve measured by two methods,
one using a rheometer (MCR 302, Anton Paar) and the other using a hand-made
stretching apparatus based on the force gauge and slider system explained in
the text. In the former method, we use small samples of width 10 mm and of
height $L=18$ mm. In the latter method, we use larger samples of width 150
mm and of height 150 mm. The sample height $L$ affects the strain rate,
which is in the range 0.025 to 0.2 s$^{-1}$ in the velocity jump experiment.
In each of plot in Fig.~\ref{Fig4} (a) to (d), there are two curves of the
same color, which correspond to the data obtained under the same condition
but with different sample and thus demonstrate the sample dependence of the
results.

In Fig.~\ref{Fig4} (a), we show results obtained with the rheometer, where
the strain rate (= 0.002 s$^{-1}$) is rather small compared with those in
the jump experiment. The results suggest (1) good reproducibility, (2) small
dependence on the stretching direction, and (3) reasonable linearity up to
the critical strain $\varepsilon_{c}\sim0.1$ although strictly linear region
is limited in the region $\varepsilon<0.02$. The stress and strain at the
breakage tend to be slightly small in the MD direction, which is barely
consistent with the fact that molecules tend to be oriented in the TD
direction. It is possible that the breakage stress could be clearly larger
than that in the TD direction, if the experiment were performed under a
better control of the sharpness of the free edge (see the next paragraph).

In Fig.~\ref{Fig4} (b), we compare the results obtained by the two methods.
The agreement of the results is good in the region $\varepsilon <\varepsilon
_{c}$, but the strain and stress at the breakage is significantly larger for
larger samples. This difference may be due to the quality of cutting the
sample. The free edge of large samples is cut in the factory at the
fabrication in the company, while the edge of smaller samples is cut in our
laboratory. The sharpness of the edge is lower for smaller samples used for
the rheometer, which leads to smaller strains at the breakage.

In Fig.~\ref{Fig4} (c) and (d), we show results obtained with the hand-made
setup, which show the dependence of the curve on the stretching speed, where
the strain rate is in the range 0.002 to 0.07 s$^{-1}$. Although the strain
rate tends to be small compared with that in the jump experiment, we see the
sample tends to become hard as the rate increases, which is consistent with
rheological measurements (see below).

\subsection{Rheological properties}

In Fig.~\ref{Fig5}, we show results of rheological measurements performed by
the rheometer (MCR 302, Anton Paar) at the frequency $f=1$ and 30 Hz.
Although PVDC is a crystalline polymer from the monomer structure, the
degree of crystallization of the sample may be relatively low and thus
transparent. As a result, we see the first relaxation of amorphous structure
and the second of crystalline structure in the plot, which respectively
correspond to the first and second peak of $\tan\delta$. \ Judging from the
first peak of $\tan\delta $, the glass transition occurs at $T\simeq$ $13$ $%
^{\circ}$C and $f=1$ Hz and at $T\simeq$ $22$ $^{\circ}$C and $f=30$ Hz. We
conclude that the characteristic time scale for the glass transition is $%
\tau\simeq1/f\simeq$ tens of milliseconds at the room temperature ($T\simeq$ 
$25$ $^{\circ}$C). The wavy feature observed in the plots of $%
E^{\prime\prime}$ and $\tan\delta$ in the region $T=-50$ to $-15$ $^{\circ}$%
C at $f=1$ Hz is due to insufficient torque and the sharp drop of $E^{\prime}
$, $E^{\prime\prime}$, and $\tan \delta$ is due to inertia. These technical
artifacts could be ameliorated but, practically, the conclusion on the
characteristic time-scale of the glass transition would not be changed.

\end{document}